\documentclass{article}%
\usepackage{amsfonts}
\usepackage{amsmath}
\usepackage{amssymb}
\usepackage{graphicx}%
\providecommand{\U}[1]{\protect\rule{.1in}{.1in}}

\begin{document}

\title{Influence of an Electric Field on the Propagation of a Photon \\in a Magnetic field }
\author{V.M. Katkov\\Budker Institute of Nuclear Physics,\\Novosibirsk, 630090, Russia\\e-mail: katkov@inp.nsk.su}
\maketitle

\begin{abstract}
In this work, a constant and uniform magnetic field is less than the Schwinger
critical value. In turn, an additional constant and uniform electric field is
taken much smaller than the magnetic field value. The propagation of a photon
in this electromagnetic field is investigating. In particular, in the presence
of a weak electric field, the root divergence is absent in the photon
effective mass near the thresholds of pair creation. The effective mass of a
real photon with a preset polarization is considered in the quantum energy
region as well as in the quasiclassical one.

\end{abstract}

\section{Introduction}

The photon propagation in electromagnetic fields and the dispersive properties
of the space region with magnetic fields is of very much interest. This
propagation is accompanied by the photon conversion into a pair of charged
particles when the transverse photon momentum is larger than the process
threshold value $k_{\perp}>2m$ (the system of units $\hbar=c=1$ is used). In
1971 Adler \cite{[1]} had calculated the photon polarization operator in a
magnetic field using the proper-time technique developed by Schwinger
\cite{[2]} and Batalin and Shabad \cite{[3]} had calculated this operator in
an electromagnetic field using the Green function found by Schwinger
\cite{[2]}. In 1971 Batalin and Shabad \cite{[3]} had calculated polarization
operator in an electromagnetic field using the Green function found by
Schwinger \cite{[2]}. In 1975 the contribution of charged-particles loop in an
electromagnetic field with $n$ external photon lines had been calculated in
\cite{[4]}. For $n=2$ the explicit expressions for the contribution of scalar
and spinor particles to the polarization operator of photon were given in
\cite{[4]}. For the contribution of spinor particles obtained expressions
coincide with the result of \cite{[3]}, but another form is used.

In this work, a constant and uniform magnetic field is less than the Schwinger
critical value $H_{0}=m^{2}/e\ =$ $4,41\cdot10^{13}$ $%
\operatorname{G}%
$. In turn, an additional constant and uniform electric field is taken much
smaller than the magnetic field value. In these fields , we consider the
polarization operator on mass shell ( $k^{2}=0,$ the metric $ab=$ $a^{0}%
b^{0}-\mathbf{ab}$ is used ) at arbitrary value of the photon energy $\omega$.

\section{General expressions}

\ \ Our analysis is based on the general expression for the contribution of
spinor particles to the polarization operator obtained in a diagonal form in
\cite{[4]} (see Eqs. (3.19), (3.33)). The eigenvalue $\kappa_{i}$ of this
operator on the mass shell $(k^{2}=0)$ determines the effective mass of the
real photon with the polarization $e_{i}$ directed along the corresponding
eigenvector:%
\begin{align}
\Pi^{\mu\nu}  &  =-\sum_{i=2,3}\kappa_{i}\beta_{i}^{\mu}\beta_{i}^{\nu
},\ \ \ \beta_{i}\beta_{j}=-\ \delta_{ij},\ \ \ \beta_{i}k=0;\ \ \label{1}\\
e_{i}^{\mu}  &  =\frac{b_{i}^{\mu}}{\sqrt{-b_{i}^{2}}},~\ b_{2}^{\mu
}=\ \left(  Bk\right)  ^{\mu}+\frac{2\Omega_{4}}{\Omega}\left(  Ck\right)
^{\mu},\ \label{2}%
\end{align}

\[
\ b_{3}^{\mu}=\left(  Ck\right)  ^{\mu}-\frac{2\Omega_{4}}{\Omega}\left(
Bk\right)  ^{\mu};
\]

\begin{align}
\mathrm{\ }\kappa_{2}  &  =r\left(  \Omega_{2}-\frac{2\Omega_{4}^{2}}{\Omega
}\right)  ,\ \ \ \ \kappa_{3}=r\left(  \Omega_{3}+\frac{2\Omega_{4}^{2}%
}{\Omega}\right)  ,\ \ \label{3}\\
\ \Omega &  =\Omega_{3}-\Omega_{2}+\sqrt{(\Omega_{3}-\Omega_{2})^{2}%
+4\Omega_{4}^{2}},\ \ r=\frac{\omega^{2}-k_{3}^{2}}{4m^{2}}.\nonumber
\end{align}
The consideration realizes in the frame where electric $\mathbf{E}$ and
magnetic\textbf{ }$\mathbf{H}$ fields are parallel and directed along the axis
3. In this frame the tensor of electromagnetic field $F_{\mu\nu}$ and tensors
$F_{\mu\nu}^{\ast}$, $B_{\mu\nu}$ and $C_{\mu\nu}$ have a form%

\begin{align}
F_{\mu\nu}  &  =C_{\mu\nu}E+B_{\mu\nu}H,\ \ F_{\mu\nu}^{\ast}=C_{\mu\nu
}H-B_{\mu\nu}E,\ \ C_{\mu\nu}=g_{\mu}^{0}g_{\nu}^{3}-g_{\mu}^{3}g_{\nu}%
^{0},\ \nonumber\\
\ B_{\mu\nu}  &  =g_{\mu}^{2}g_{\nu}^{1}-g_{\mu}^{1}g_{\nu}^{2};\ \ eE/m^{2}%
=E/E_{0}\equiv\nu,\ \ eH/m^{2}=H/H_{0}\equiv\mu;\label{eq3}\\
\Omega_{i}  &  =\frac{\alpha m^{2}}{\pi}\mu\int\limits_{-1}^{1}\ dv\int
\limits_{0}^{\infty\text{ }}\ f_{i}(v,x)\exp(\text{\textrm{i}}\psi(v,x))dx.
\label{eq4}%
\end{align}
Here%

\begin{align}
\frac{1}{\nu x}f_{1}  &  =\frac{\cos(\mu xv)\cosh(\nu xv)}{\sin(\mu
x)\sinh(\nu x)}-\frac{\cos(\mu x)\cosh(\nu x)\sin(\mu xv)\sinh(\nu xv)}%
{\sin^{2}(\mu x)\sinh^{2}(\nu x)},\nonumber\\
\frac{1}{\nu x}f_{2}  &  =2\frac{\cosh(\nu x)(\cos(\mu x)-\cos(\mu xv))}%
{\sinh(\nu x)\sin^{3}(\mu x)}+f_{1},\ \nonumber\\
\frac{1}{\nu x}\ \ f_{3}  &  =2\frac{\cos(\mu x)(\cosh(\nu x)-\cosh(\nu
xv))}{\sin(\mu x)\sinh^{3}(\nu x)}-f_{1},\nonumber\\
\frac{1}{\nu x}f_{4}  &  =\frac{\cos(\mu x)\cos(\mu xv)-1}{\sin^{2}(\mu
x)}\frac{\cosh(\nu x)\cosh(\nu xv)-1}{\sinh^{2}(\nu x)}\nonumber\\
&  +\frac{\sin(\mu xv)\sinh(\nu xv)}{\sin(\mu x)\sinh(\nu x)};\label{eq5}\\
\ \ \psi(v,x)  &  =2r\left(  \frac{\cosh(\nu x)-\cosh(\nu xv)}{\nu\sinh(\nu
x)}+\frac{\cos(\mu x)-\cos(\mu xv)}{\mu\sin(\mu x)}\right)  -x. \label{eq6}%
\end{align}
Let us note that the integration contour in Eq.(\ref{eq4}) is passing slightly
below the real axis.

After all calculations have been fulfilled we can return to a covariant form
of the process description using the following expressions%

\begin{align}
E^{2},H^{2}  &  =\left(  \mathcal{F}^{2}+\mathcal{G}^{2}\right)  ^{1/2}%
\pm\mathcal{F,\ \ F=}\left(  \mathbf{E}^{2}-\mathbf{H}^{2}\right)
\diagup2,\ \ \mathcal{G=}\mathbf{EH},\ \nonumber\\
\left(  C^{2}\right)  _{\mu\nu}  &  =\left(  F_{\mu\nu}^{2}+H^{2}g_{\mu\nu
}\right)  \diagup\left(  E^{2}+H^{2}\right)  ,\ \ \left(  C^{2}\right)
_{\mu\nu}-\left(  B^{2}\right)  _{\mu\nu}=g_{\mu\nu}. \label{eq7}%
\end{align}

The real part of $\kappa_{i}$ determines the refractive index $n_{i}$ of
photon with the polarization $e_{i}^{\mu}=\beta_{i}^{\mu},$ $i=1,2$:%

\begin{equation}
\ \ \kappa_{i}=(m_{i}^{ef})^{2},\ \ n_{i}=1-\frac{\mathrm{\operatorname{Re}%
}\kappa_{i}}{2\omega^{2}}. \label{4}%
\end{equation}
At $r>1,$ the proper value of polarization operator $\kappa_{i}$ includes the
imaginary part, which determines the probability per unit length of pair production:%

\begin{equation}
W_{i}=-\frac{1}{\omega}\mathrm{\operatorname{Im}}\kappa_{i}. \label{5}%
\end{equation}

Let us not that in considered case, we have $\mu\ll1,\ \nu\ll\mu.$ The high
energy region, $r\gtrsim1/\mu^{2},$ is contained in the region of the standard
quasiclassical approximation (SQA) \cite{[5]}. In SQA, the effective mass of a
photon depends on the parameter $\kappa$ only: $\kappa^{2}=4r(\mu^{2}+\nu
^{2})=-(Fk)^{2}/m^{2}H_{0}^{2}.$For $\kappa\gtrsim1$ the influence of weak
electric field on the polarization operator is small. Because of this we
consider now the case of energies $r\ll1/\mu^{2}.$

\section{Region of intermediate photon energies}

Choose a point $x_{0}$ in the following way: $\varphi^{\prime}(x_{0})=0$, were%

\begin{equation}
\varphi\left(  x\right)  =-\psi(0,x)=2r\left(  \frac{1}{\mu}\tan\frac{\mu
x}{2}-\frac{1}{\nu}\tanh\frac{\nu x}{2}\right)  +x.\ \label{6}%
\end{equation}
Then, we have the following equation for $x_{0}$:%

\begin{equation}
\tan^{2}\frac{\nu s}{2}+\tanh^{2}\frac{\mu s}{2}=\frac{1}{r},\ \ x_{0}%
=-\mathrm{i}s. \label{06}%
\end{equation}
We represent the integral for $\Omega_{i}$ as%

\begin{equation}
\Omega_{i}=\frac{\alpha m^{2}}{\pi}(a_{i}+b_{i}), \label{7}%
\end{equation}
where%

\begin{align}
a_{i}  &  =\mu%
{\textstyle\int\limits_{-1}^{1}}
dv%
{\displaystyle\int\limits_{0}^{x_{0}}}
f_{i}(v,x)\exp[\mathrm{i}\psi(v,x)]dx,\label{71}\\
b_{i}  &  =\mu%
{\textstyle\int\limits_{-1}^{1}}
dv%
{\textstyle\int\limits_{x_{0}}^{\infty}}
f_{i}(v,x)\exp[\mathrm{i}\psi(v,x)]dx. \label{72}%
\end{align}
In the integral $a_{i}$ in Eq. (\ref{7}), the small values $x\sim1$
contribute. We calculate this integral expanding the entering functions over
$x,$ take into account that in the region under consideration, the condition
$r\mu^{2}\ll1$ is fulfilled. Then in exponent, we keep the term $-x$ only and
extend the integration over $x$ to infinity. In the result of not complicated
integration over $v,$ we have:%

\begin{align}
a_{2}  &  =-\frac{16}{45}(\mu^{2}+\nu^{2}),\ \ \ a_{3}=-\frac{28}{45}(\mu
^{2}+\nu^{2}),\ \ \nonumber\\
\ \kappa_{2}^{a}  &  =-\frac{4\alpha m^{2}\kappa^{2}}{45\pi},\ \ \ \kappa
_{3}^{a}=-\frac{7\alpha m^{2}\kappa^{2}}{45\pi},\ \ \ \ \kappa^{2}%
=-\frac{(Fk)^{2}}{H_{0}^{2}m^{2}}. \label{8}%
\end{align}
These asymptotics are well known.

In the integral $b_{i}$ Eq. (\ref{72}), small values $v$ contribute. Expanding
entering functions over $v$ and extending the integration over $v$ to
infinity, we get%

\begin{align}
b_{i}  &  =\mu\int\limits_{-\infty}^{\infty}\ dv\int\limits_{x_{0}}%
^{\infty\text{ }}\ f_{i}(0,x)\exp\left\{  -\text{\textrm{i}}\left[
\varphi\left(  x\right)  +v^{2}\chi\left(  x\right)  \right]  \right\}
dx,\label{9}\\
\chi\left(  x\right)   &  =rx^{2}\left(  \frac{\nu}{\sinh(\nu x)}-\frac{\mu
}{\sin(\mu x)}\right)  . \label{09}%
\end{align}
After the integration over $v,$ one has%

\begin{equation}
b_{i}=\mu\sqrt{\pi}\exp\left(  -\mathrm{i}\frac{\pi}{4}\right)
{\displaystyle\int\limits_{x_{0}}^{\infty}}
\frac{f_{i}(0,x)}{\sqrt{\chi(x)}}\exp\left[  -\mathrm{i}\varphi(x)\right]  dx.
\label{009}%
\end{equation}

The case $\mu^{-2}\gg r\gg\mu^{-2/5}$ was considered in \cite{[6]} (see Eq.
31). The result has the following invariant form:%

\begin{align}
\kappa_{2}  &  =\frac{\alpha m^{2}\kappa}{8}\sqrt{\frac{3}{2}}\exp\left(
-\frac{8}{3\kappa}+\frac{64\widetilde{\mathcal{F}}}{15\kappa^{3}}\right)
,\ \ W_{3}=2W_{2},\label{k1}\\
\kappa &  =4r\sqrt{\mu^{2}+\nu^{2}},\ \ \ \widetilde{\mathcal{F}}=\frac
{\nu^{2}-\mu^{2}}{2}. \label{k2}%
\end{align}

Leaving the main terms of the expansion in the parameter $\xi=\nu/\mu\ll1,$ we have%

\begin{equation}
\kappa_{2}\left(  \xi\right)  /\kappa_{2}\left(  0\right)  =\exp\left[
\frac{2\Delta}{3\sqrt{r}}\left(  1+\frac{1}{r}\right)  \right]  . \label{k3}%
\end{equation}

\section{The quantum energy region near the lower thresholds}

We consider now the energy region where $r-1\ \ll1$ when the moving of created
particles is nonrelativistic. In this case, Eq.(\ref{06}) and its solutions
are given by the following approximate equations%
\begin{align}
\frac{\xi^{2}l^{2}}{16}  &  \simeq\exp(-l)+\frac{1-r}{4},\ \ l=\mu
s,\ \xi=\frac{\nu}{\mu};\label{91}\\
\ l  &  \simeq2\ln\frac{4}{\xi\ln\frac{4}{\xi}},\ \ |r-1|\ \ll\xi^{2}%
l^{2};\label{92}\\
l  &  \simeq\ln\frac{4}{r-1},\ \ r-1\gg\xi^{2}l^{2}; \label{93}%
\end{align}
Leaving the main terms of the expansion, we obtain%

\begin{align}
\mathrm{i}\varphi(x)  &  \simeq\beta(r)-\gamma e^{-\mathrm{i}z}-\mathrm{i}%
zq+\mathrm{i}\Delta x^{3}/12,\label{101}\\
\ x  &  =(z-\mathrm{i}l)/\mu,\ \ \ q=(r-1)/\mu,\ \Delta=\frac{\xi^{2}}{\mu
},\label{102}\\
\ \beta(r)  &  =\frac{2r}{\mu}-ql(r),\ \gamma=q+\Delta l^{2}/4,\ \ \label{103}%
\\
\ \chi(x)  &  \simeq x,\ \ \ f_{1,2,4}(0,x)\simeq0,\ \ \ f_{3}(0,x)\simeq
-\mathrm{i.} \label{104}%
\end{align}
In the region of lower thresholds, where the particles occupy not very high
energy levels, we present Eq. (\ref{009}) for $b_{3}$ in the form%

\begin{align}
b_{3}  &  =-\mathrm{i}\sqrt{\pi\mu}\exp\left(  -\mathrm{i}\frac{\pi}%
{4}\right)  \exp\left[  -\beta(r)\right] \nonumber\\
&  \cdot%
{\textstyle\int\limits_{0}^{\infty}}
\frac{dz}{\sqrt{z-\mathrm{i}l}}%
{\textstyle\sum\limits_{k=0}^{\infty}}
\frac{\gamma^{k}}{k!}\exp\{\mathrm{i}[(q-k)z-\Delta(z-\mathrm{i}%
l)^{3}/12]\},\ \label{10}%
\end{align}
If $|\delta|\ll1,\ \delta=q-n,\ \ \Delta\ll1,\ \gamma\simeq n,$ the large $z$
contributes and we have after the change of variables%

\begin{align}
b_{3}  &  \simeq-\mathrm{i}\exp\left(  -\mathrm{i}\frac{\pi}{4}\right)
\sqrt{\mu\pi}\exp[-\beta(r)]\nonumber\\
&  \cdot\frac{n^{n}}{n!}%
{\textstyle\int\limits_{0}^{\infty}}
\frac{dz}{\sqrt{z}}\exp[\mathrm{i}(\delta z-\Delta z^{3}/12)]. \label{11}%
\end{align}
After integration over $z,$ we have the following approximate expressions%

\begin{equation}
b_{3}=2\sqrt{\mu\pi}\frac{n^{n}}{n!}\exp[-\beta(r)]d(\delta,\Delta
),\ \ \ \ \ \ \ \ \ \ \label{111}%
\end{equation}
were%

\begin{align}
d(\delta,\Delta)  &  =-\exp\left[  \mathrm{i}\frac{\pi}{2}\vartheta
(\delta)\right]  \sqrt{\frac{\pi}{\delta}},\ \ \ \ \ \ \Delta\ll|\delta
|^{3};\label{112}\\
d(\delta,\Delta)  &  =-\frac{1}{6}\exp\left(  \mathrm{i}\frac{\pi}{6}\right)
\Gamma\left(  \frac{1}{6}\right)  \left(  \frac{12}{\Delta}\right)
^{1/6}\ ,\ \ \ |\delta|\ll\Delta^{1/3} \label{113}%
\end{align}

\begin{equation}
\ \ \label{12}%
\end{equation}
where $\vartheta(z)$ is Heaviside function: $\vartheta(z)=1$ for
$z\geqslant0,$ $\vartheta(z)=0$ for $z<0.$ The expression for $\kappa_{3}$
with the accepted accuracy can be rewritten in the following form:%

\begin{align}
\kappa_{3}^{b}  &  \simeq-\alpha m^{2}\sqrt{\frac{\mu}{|\delta|}}\exp\left[
\mathrm{i}\frac{\pi}{2}\vartheta(\delta)\right]  e^{-\zeta}\frac{(2\zeta)^{n}%
}{n!},\ \ \Delta\ll|\delta|^{3},\ \zeta=2r/\mu.\label{13}\\
\kappa_{3}^{b}  &  \simeq-\frac{\sqrt{3}+\mathrm{i}}{12}\alpha m^{2}%
\sqrt{\frac{\mu}{\pi}}\Gamma\left(  \frac{1}{6}\right)  \left(  \frac
{12}{\Delta}\right)  ^{1/6}e^{-\zeta}\frac{(2\zeta)^{n}}{n!},\ \ \ |\delta
|\ll\Delta^{1/3},\ \label{14}%
\end{align}
In considered quantum case $n$ $\sim1$ the real part of $\kappa_{3}^{b}$ is
exponentially small compared to $\kappa_{3}^{a}$ (\ref{8}), but the imaginary
part of the effective mass is given by Eq. (\ref{13}).

At $\Delta\gtrsim1,$ we have $\gamma\gg1,$ and the small $z$ contributes to
the integral in Eq. (\ref{009})$.$ In this case, it is possible to use the
method of stationary phase%

\begin{equation}
\kappa_{3}^{b}\simeq-\mathrm{i}\alpha m^{2}\sqrt{\frac{\mu}{2\left(  q+\Delta
l^{2}/4\right)  l}}\exp[-\beta(r)+\gamma+\Delta l^{3}/12] \label{140}%
\end{equation}

\section{Low energy region $r-1<0$}

The case $1-r\ll$ $\nu^{2/3}$ was discussed above(see Eq. (\ref{14})) where we
have to put $n=0$. For the case $1\gg$ $1-r\gg\nu^{2/3},$ the following expression%

\begin{equation}
l\simeq\frac{2}{\xi}\sqrt{1-r},\ \ \label{150}%
\end{equation}
is valid and one can use the method of stationary phase. Then we have%

\begin{equation}
\kappa_{3}^{b}\simeq-\mathrm{i}\frac{\alpha m^{2}\mu}{\sqrt{1-r}}\exp\left[
-\frac{2r}{\mu}-\frac{4\left(  1-r\right)  ^{3/2}}{3\nu}\right]  \label{15}%
\end{equation}
$\ $

This method is valid for lower photon energy when $\nu^{-2}\gg1/r-1\gg
\nu^{2/3}.$ Then the approximate solution of Eq. (\ref{06}) has form%

\begin{align}
s  &  \simeq\frac{2}{\nu}\arctan\sqrt{\frac{1}{r}-1},\ \ \sqrt{-\frac
{\chi\left(  x_{0}\right)  \varphi^{\prime\prime}\left(  x_{0}\right)  }%
{2}\simeq}\frac{\nu s}{2},~\nonumber\\
\ f_{3}\left(  x_{0}\right)   &  \simeq\mathrm{i}\frac{\nu s}{r\sqrt{r\left(
1-r\right)  }},\ \ \ \ \ \ f_{1,2,4}(0,x)\simeq
0,\ \ \ \ \ \ \ \ \ \ \ \ \ \ \ \ \ \label{160}\\
\kappa_{3}^{b}  &  \simeq-\mathrm{i}\frac{\alpha m^{2}\mu}{\sqrt{r(1-r)}}%
\exp\left[  -\frac{2r}{\mu}-\frac{2}{\nu}\arctan\sqrt{\frac{1}{r}-1}+\frac
{2r}{\nu}\sqrt{\frac{1}{r}-1}\right]  . \label{16}%
\end{align}

For $1-r\ll1,$ expanding the incoming in Eq. (\ref{16}) functions we have Eq.
(\ref{15}). When the value $r$ small enough $\nu^{3}\ll r^{3/2}\ll\nu,$
leaving the leading term of decomposition over $r,$ we get%

\begin{equation}
\kappa_{3}^{b}\simeq-\mathrm{i}\frac{\alpha m^{2}\mu}{\sqrt{r}}\exp\left[
-\frac{2r}{\mu}-\frac{\pi}{\nu}+\frac{4\sqrt{r}}{\nu}\right]  . \label{17}%
\end{equation}
The imaginary part of this expression coincides with Eq. (40) \cite{[6]}. The
ratio of this expression to the corresponding formula in a pure electric field
has the form%

\begin{equation}
\kappa_{3}^{b}/\kappa_{3}^{b}\left(  \mu=0\right)  =\frac{\pi}{\xi}\exp\left[
r\left(  \frac{\pi}{\nu}-\frac{2}{\mu}\right)  \right]  . \label{18}%
\end{equation}
Eqs. (\ref{17}), (\ref{18}) are valid for the photon energy $r\gg\nu^{2}.$ For
the region of the photon energy $r\lesssim\nu^{2},$ we can use the results of
\cite{[6]}, where we must put $\eta=1/\xi.$%

\begin{align}
W_{3}  &  =\frac{2\pi}{\xi}\alpha m^{2}\exp\left(  -\frac{\pi}{\nu}\right)
I_{1}^{2}\left(  z\right)  ,\ \ \ \ \ W_{2}\ll W_{3},\label{19}\\
\ z  &  =2\frac{\sqrt{r}}{\nu},\ \ \ \ \ \ W_{3}/W_{3}\left(  \mu=0\right)
=\pi/\xi. \label{20}%
\end{align}
For $r\gg\nu^{2}$ the asymptotic representation $\mathrm{I}_{n}\left(
z\right)  \simeq\exp\left(  z\right)  \diagup\sqrt{2\pi z}$ can be used. As a
result one obtains Eq.(\ref{17}) where the leading terms have to be retained.
At very low photon energy $r\ll\nu^{2},$ using the expansion of the Bessel
functions for the small value of argument, we have%

\begin{equation}
W_{3}=\frac{2\pi}{\xi}\frac{\alpha m^{2}r}{\omega\nu^{2}}\exp\left(
-\frac{\pi}{\nu}\right)  . \label{21}%
\end{equation}

\section{Conclusion}

We have considered the polarization operator of a photon in a constant
magnetic fields in the presence of a weak electric field. The effective mass
of a photon was calculated using three different overlapping approximation. In
the region of SQA applicability, the created by a photon particles have
ultrarelativistic energies. The role of fields in this case is to transfer the
required transverse momentum and the electric field actions less than that of
the magnetic field. At lower energies, the role of the electric field
increases. It is necessary to note a special significance of a weak electric
field $E=\xi H$ $(\xi\ll1)$ in the removal of the root divergence of the
probability when the particles of pair are created on the Landau levels with
the electron and positron momentum $p_{3}=0$ \cite{[5]}. The frame is used
where $k_{3}=0.$

Generally speaking, at $\xi\ll1$ the formation time $t_{c}$ of the process
under consideration is $1/\mu.$ Here we use units $\hbar=c=m=1.$ At this time
the particles of creating pair gets the momentum $\delta p_{3}\sim\xi.$ If the
value $\xi^{2}$ becomes larger than the distance apart Landau levels $\mu$
$(\Delta=\xi^{2}/\mu\gg1)$ all levels overlapped. Under this condition the
divergence of the probability is vanished and the method of stationary phase
is valid even in the energy region $r-1\lesssim\mu,$ whereas that is
inapplicable in the absence of electric field \cite{[5]}. In the opposite case
$\nu^{2}\ll\mu^{3}$ for the small value of $p_{3}\ll\sqrt{\mu},$ in the region
where the influence of electric field is negligible, the formation time of the
process $t_{f}$ is $1/p_{3}^{2}$ and $\delta p_{3}\sim\nu/p_{3}^{2}$ $\ll
p_{3}$ .\ It is follows from above that $\nu^{1/3}\ll p_{3}\ll\sqrt{\mu}$ . At
this condition the value of discontinuity is $\sqrt{t_{f}/t_{c}}\sim\sqrt{\mu
}/p_{3}.$For $\nu^{1/3}\gg p_{3}$ the time $t_{f}$ is determined by the
self-consistent equation $\delta\varepsilon^{2}\sim1/t_{f}\sim\nu^{2}t_{f}%
^{2},$ $t_{f}\sim\nu^{-2/3}$ and the value of discontinuity becomes $\sqrt{\mu
t_{f}}\sim(\mu^{3}/\nu^{2})^{1/6}$ instead of $\sqrt{\mu}/p_{3}.$

In the region $\omega\lesssim2m$ $(r\lesssim1)$ the energy transfer from
electric field to the created particles becomes appreciable and for $\omega
\ll$ $m$ it determines the probability of the process mainly. At $\omega\ll
eE/m$ $(\sqrt{r}\ll\nu),$ the photon assistance in the pair creation comes to
the end and the probability under consideration defines the probability of
photon absorption by the particles created by electromagnetic fields. The
influence of a magnetic field on the process is connected with the interaction
of the magnetic moment of the created particles and magnetic field. This
interaction, in particular, has appeared in the distinction of the pair
creation probability by field for scalar and spinor particles \cite{[2]}.

The work was supported by the Ministry of Education and Science of the Russian
Federation. The author is grateful to the Russian Foundation for Basic
Research grant \#15-02-02674 for partial support of the research.

\end{document}